# Invisibility Dips of Near-Field Energy Transport in A Spoof Plasmonic Meta-Dimer

*Fei Gao, Zhen Gao, Yu Luo\*, Baile Zhang\**

F. Gao, Z. Gao, Prof. B. Zhang
Division of Physics and Applied Physics, School of Physical and Mathematical Sciences, Nanyang Technological University, Singapore 637371, Singapore.
Email: blzhang@ntu.edu.sg (B. Zhang)

Prof. Y. Luo
School of Electrical and Electronic Engineering, Nanyang Technological University, Nanyang Avenue 639798, Singapore.
Email: luoyu@ntu.edu.sg (Y. Luo)

Prof. B. Zhang
Centre for Disruptive Photonic Technologies, Nanyang Technological University, Singapore 637371, Singapore.

## Abstract

Invisibility dips, minima in scattering spectrum associated with asymmetric Fano-like line-shapes, have been predicted with transformation optics in studying strong coupling between two plasmonic nanoparticles. This feature of strongly coupled plasmonic nanoparticles holds promise for sensor cloaking. It requires an extremely narrow gap between the two nanoparticles, though, preventing its experimental observation at optical frequencies. Here, the concept of spoof surface plasmons is utilized to facilitate the strong coupling between two spoof-localized-surface-plasmon (SLSP) resonators. Instead of observing in far field, the near-field energy transport is probed through the two SLSP resonators. By virtue of enhanced coupling between the two resonators stacked vertically, a spectral transmission dip with asymmetric Fano-like line-shape, similar to the far-field "invisibility dips" predicted by transformation optics, is observed. The underlying mode interference mechanism is further demonstrated by directly imaging the field maps of interfered waves that are tightly localized around the resonators. These near-field "invisibility dips" may find use in near-field sensing, on-chip switching, filters and logical operation elements.

**1. Introduction**

The recent re-examination of strong coupling between two plasmonic nanoparticles by transformation optics revealed a new phenomenon of "invisibility dips", or minima in the scattering spectrum,[1] which exhibit asymmetric line-shapes similar to Fano resonances,[2-4] but arises from destructive mode interference of surface plasmon resonances. This mode interference is between two successive localized multipolar modes different from that between the propagating surface waves and localized surface waves[5]. A plasmonic dimer that can focus electromagnetic (EM) energy at the frequencies of invisibility dips but with zero far-field scattering cross section, shows promising potential applications in sensor cloaking that can enhance a near-field signal while invisible itself.[1] However, the original transformation-optics prediction requires an extremely narrow gap between the two nanoparticles in order to facilitate the strong coupling,[1] making it difficult to observe in practice.

As EM modes analogous to optical surface plasmons at metal/dielectric interfaces, spoof surface plasmons appear at much lower frequencies and are supported by the underlying textured metal structures.[6-11] They hold promise in microwave- to infrared-frequency device applications,[12-13] as their simple tunability of near-field coupling is desirable in novel EM functionalities.

Here, we investigate the strong coupling between two spoof-localized-surface-plasmon (SLSP) resonators stacked vertically, in contrast to most previous studies on the interaction between SLSP particles placed on a two-dimensional plane,[8,14-15] in which the near-field coupling is generally insufficient to create invisibility dips. In the vertical coupling configuration, despite a sufficient gap between the two resonators, the coupling between them can still be strong enough because of their large area overlap. Particularly, the strong interaction contains significant radiation coupling that induces lifetime-contrast mode splitting, and thus is fundamentally different from previously investigated phenomena in field enhancement and Fano resonance.[8,15] The "invisibility dips" predicted by transformation optics for the far-field scattering, can be observed in the transmission spectrum for the near-field energy transport through the two resonators.[16] Although mode splitting (also called Autler-Towners splitting) between two optical resonators is widely demonstrated,[17] few works have directly observed the mode splitting with contrast lifetimes. Furthermore, our study is the first direct imaging on the process of strong-coupling-induced near-field interference between successive multipolar modes.

## 2. Implementing Strong Coupling with SLSP Dimer

To illustrate that the larger area overlap gives rise to the stronger coupling in vertical configuration, we quantitatively compare the transmission spectrum of horizontal and vertical coupling configurations with simulation. In **Figure 1a**, b, both the hexapole and octopole modes are split in the horizontal and vertical coupling configurations. The edge-edge distance between resonators in both configurations are $D = 6.1$ mm as shown in the insets. To be consistent with later experiment, the horizontally coupled dimer (inset in Figure 1a) is placed on a Teflon plate with thickness $t = 6.1$mm, and the vertically coupled dimer (inset in Figure 1b) is spaced by the same Teflon plate. Mode splitting in the vertical coupling configuration (shown in Figure 1b) exhibits larger spectral separations (i.e. hexapole mode splitting with $\delta\omega_{Hv} = 0.412$ GHz, and octopole mode splitting with $\delta\omega_{Ov} = 0.284$ GHz, where the subscripts 'H' and 'O' represent hexapole and octopole modes respectively, 'v' denotes vertical coupling) than those in the horizontal coupling configuration shown in Figure 1a (i.e. hexapole mode splitting with $\delta\omega_{Hh} = 0.072$ GHz, and octopole mode splitting with $\delta\omega_{Oh} = 0.030$ GHz, where the subscript 'h' denotes horizontal coupling). The larger mode splitting separation implies a stronger coupling in the vertical coupling configuration. In later analysis we will use a complex number $\kappa$ to quantify the coupling strength. It is worth mentioning that while the linewidths (or lifetimes) of split modes are almost the same in the horizontal coupling configuration, they are different in the vertical coupling configuration. For instance, in Figure 1b the linewidths of split hexapole modes differ by $\delta\gamma_H = 0.0104$ GHz, and those of split octopole modes differ by $\delta\gamma_O = 0.0002$ GHz. Similar to the formation of super-radiant states in coherent spontaneous radiation,[18] this lifetime-contrast splitting can be attributed to the enhanced radiation coupling, as a result of the large area overlap between the two resonators.

## 3. Multipolar Modes in a Single SLSP Resonator

We first study the near-field properties of eigen modes in a single SLSP resonator, which consists of periodic grooves etched on an 18-μm-thick metallic disk. The periodicity and groove width are $d = 1.256$ mm and $a = 0.628$ mm, respectively. The inner and outer radii of the resonator are $r = 3.0$ mm, $R = 12.0$ mm, respectively. A 254-μm-thick Rogers RT5880 dielectric substrate which has relative permittivity of 2.2+0.0009$i$, is used to support the thin resonator. The whole structure is placed on the top of a 9.9-mm-thick Teflon plate (relative permittivity 2.1)[19] which will be later used as a separation layer in the vertical coupling configuration. Similar to previous works,[11, 20] the near-field excitation is achieved with a monopole antenna at point 'S' (shown in **Figure 2a**), and the transmission spectrum is detected with another monopole antenna at point 'P' (shown in Figure 2a). For later observation of mode interference, the two modes we choose are hexapole at 5.258 GHz and octopole at 5.599 GHz, whose field patterns are measured as shown in Figure 2b, c, respectively.

## 4. Theoretical Analysis with Coupled Mode Theory

We then attach a second resonator on the backside of the Teflon plate to fulfil the vertical coupling configuration. Teflon plates with two different thicknesses $t = 9.9$ mm and $t = 8.2$ mm are adopted to tune the coupling between the two resonators. First, simulation results in **Figure 3a** show that both hexapole (resonance frequency $\omega_H =$ 5.305 GHz, linewidth $\gamma_H = 0.008$ GHz) and octopole (resonance frequency $\omega_o = 5.725$ GHz, linewidth $\gamma_o = 0.0027$ GHz) modes of an individual resonator are split into two modes with different linewidths. The lower and higher peaks are named as binding and anti-binding modes indicated with subscript 'b' and 'a' respectively. This lifetime-contrast splitting can be understood with coupled mode theory (CMT).[21] Equations of coupled modes are set up as follows:

$$\begin{cases} i\omega \begin{bmatrix} H_1 \\ O_1 \end{bmatrix} = \begin{bmatrix} i\omega_H - \gamma_H & 0 \\ 0 & i\omega_O - \gamma_O \end{bmatrix} \begin{bmatrix} H_1 \\ O_1 \end{bmatrix} + \begin{bmatrix} i\kappa_H & 0 \\ 0 & i\kappa_O \end{bmatrix} \begin{bmatrix} H_2 \\ O_2 \end{bmatrix} + \begin{bmatrix} \tau_H \\ \tau_O \end{bmatrix} S_{in} \\ i\omega \begin{bmatrix} H_2 \\ O_2 \end{bmatrix} = \begin{bmatrix} i\omega_H - \gamma_H & 0 \\ 0 & i\omega_O - \gamma_O \end{bmatrix} \begin{bmatrix} H_2 \\ O_2 \end{bmatrix} + \begin{bmatrix} i\kappa_H & 0 \\ 0 & i\kappa_O \end{bmatrix} \begin{bmatrix} H_1 \\ O_1 \end{bmatrix} \\ S_{out} = \cos(m_H \varphi) \eta_H H_2 + \cos(m_O \varphi) \eta_O O_2 \end{cases} \quad (1)$$

where '$H$' and '$O$' refer to hexapole and octopole modes respectively, subscript '1' and '2' denote different resonators, and symbols '$\omega$', '$\gamma$', '$\tau$' and '$\eta$' represent resonance frequency, dissipation loss, input coupling strength and output coupling strength respectively. Since the monopole probe is placed at the opposite position of the source, the phase difference is $\varphi = \pi$. Mode orders are $m_H = 3$ for hexapole mode, and $m_O = 4$ for octopole mode, respectively. The measured transmission spectra are defined as $T = |\frac{S_{out}}{S_{in}}|$. Through CMT fitting, the complex coupling strength can be retrieved as $\kappa_H = 0.0762+0.005i$ (GHz) for the hexapole mode and $\kappa_o = 0.036+0.0001i$ (GHz) for the octopole mode. Other parameters can be obtained as $\tau_H = \eta_H = 0.8422$, and $\tau_O = \eta_O = 0.9911$ for the input and output couplings. Real parts of $\kappa$ are contributed by both evanescent field and radiation coupling, while imaginary parts of $\kappa$ are only induced by radiation coupling,[22-24] which can lead to split modes with contrast lifetimes similar to the formation of superradiant modes in coherent spontaneous radiation.[18] Here the radiation coupling is nontrivial, as a result of large area overlap between the two resonators with relatively small separation. One resonator can thus receive almost half of the radiation from the other. Experimentally detected transmission in Figure 3b verify the lifetime-contrast splitting, in which binding modes show smaller line width than antibinding modes.

## 5. Experimental Demonstration of Invisibility Dips

In order to gain further insight into the mode splitting, field profiles of the binding mode $O_b$ at 5.58 GHz and the anti-binding mode $O_a$ at 5.62 GHz in the X-Y plane are measured by the near-field scanning system with a spacer thickness $t = 9.9$ mm (profiles of the hexapole mode can be found in Ref. [20]). Observed field patterns 0.5mm above/below the resonator dimer are shown in Figure 3c and Figure 3d. For further illustration, we show simulated field patterns of $O_b$ (Figure 3e) and $O_a$ (Figure 3f) in the cross section of the resonator dimer in the X-Z plane. Inside the Teflon layer in Figure 3e, the $E_z$ field points from the top resonator to the bottom one, indicating that the two resonators are bound by opposite charge distributions. However inside the spacer in Figure 3f, the $E_z$ field points from both top and bottom resonators to the middle region, demonstrating that the two resonators are anti-bond and repulsed by the same charge distributions.

Next, as we decrease the Teflon thickness to $t = 8.2$ mm, binding and anti-binding modes of both hexapole and octopole are spectrally separated further in Figure 3a. Fitting with Eq. (1), retrieved coupling strength are $\kappa_H = 0.1212+0.0052i$ (GHz) for the hexapole mode and $\kappa_o = 0.0702+0.0001i$ (GHz) for the octopole mode, in which the real parts are much larger than their counterparts when $t = 9.9$ mm, while imaginary parts are almost unchanged. The reason is that a closer separation between resonators can enhance their near-field coupling, as reflected in the real part of coupling strength. Yet as one resonator has already covered almost half space from the viewpoint of the other resonator, the radiation coupling between them is almost unaffected by the further separation reduction. The experimentally measured transmission spectrum is shown in Figure 3b, which show good agreements with simulations and CMT results. In addition, it can be found that decreasing the spacer thickness shifts the anti-binding hexapole mode $H'_a$ and the binding octopole mode $O'_b$ close to each other. This shows the possibility for these two modes to spectrally overlap and interfere with each other. This spectral overlap requires extremely strong coupling between the two SLSP resonantors, which is difficult to attain using the transverse arrangement in the horizontal coupling configuration in previous experiments.[15] As we will show later, this spectral overlap enables us to observe the effect of invisibility dip predicted by transformation optics.[1]

To achieve the mode interference, we further decrease Teflon's thickness to $t = 6.1$ mm. The transmission spectrum is shown in **Figure 4a**. $H_a$ and $O_b$ are closer than in Figure 3a. They are no longer in symmetric Lorentz lineshape. An invisibility dip forms between the resonant frequencies of the two modes which is consistent with the theoretical prediction.[1] By fitting the simulation curve with Eq. (1), the coupling strengths are retrieved as $\kappa_H = 0.2080+0.0048i$ (GHz), $\kappa_o = 0.1420+0.0001i$ (GHz). The experimentally detected asymmetric lineshape in Figure 4c verifies the simulation and CMT results.

To further illustrate how anti-binding ($H_a$) and binding ($O_b$) modes interact with each other, simulated mode profiles at three frequencies (i.e. $H_a$, Dip, and $O_b$ in Figure 4a) are shown in Figure 4b. The phase difference between top and bottom mode profiles of $O_b$ and $H_a$ are similar with those in Figure 3c,d, and are verified by experimental results shown in Figure 4d. In addition, we also show the simulated patterns on the middle plane inside the Teflon spacer as in Figure 4b. As discussed before, the field pattern on the middle plane of anti-binding mode $H_a$ is almost not discernable, but that of the binding mode $O_b$ is comparable with those on top and bottom planes. The interference of $H_a$, and $O_b$ is depicted at the dip frequency in the middle panel of Figure 4b. Two features can be noticed: first, at the probe position, anti-binding $H_a$ and binding $O_b$ modes destructively interfere with each other, leading to the null field; second, in the middle plane, octopole mode dominates while the hexapole mode is relatively weak. This shows that the null field is indeed caused by the mode interference between $H_a$ and $O_b$. Experiment results in Figure 4d confirm the appearance of mode interference on the top and bottom surface of the resonator dimer.

Then, we further demonstrate that transmission dips with asymmetry line shapes arise from mode interference rather than Fano interaction. The asymmetry property of the lineshape can be flipped by further decreasing the spacer thickness to $t = 4.9$ mm. The underlying mechanism is mode crossing that causes the frequency of the hexapole anti-binding mode to be higher than that of the octopole binding mode as shown in **Figure 5a**. This phenomenon has been further verified with measured near-field patterns shown in Figure 5b. Following the above fitting process, the retrieved coupling strengths are $\kappa_H = 0.265+0.0049i$ (GHz), $\kappa_o = 0.2+0.0001i$ (GHz). Note that anomalous modes, arising from mode coupling are not observed here.[25] Finally, decreasing the spacer thickness to $t = 2.1$ mm, we observe that the binding mode ($H_b$) of hexapole shifts to a lower frequency (4.916 GHz in Figure 5c) than that of above samples. That means the coupling in this scenario is much stronger than before. Associated with the shift of $H_b$, a transmission dip at 4.95 GHz (in Figure 5c) is observed. Using near-field scanning technology, it has been verified that this dip originates from the interference between the binding mode ($H_b$) of hexapole and the antibinding mode ($Q_a$) of quadrapole (shown in Figure 5d).

**6. Sensing with invisibility dips**

The invisibility dips, arising from the interference of multipolar modes, are sensitive to the change of surrounding materials. That is because the surrounding materials can change the interference condition of multipolar modes. Here, the meta-dimer with 2.1-mm-thick Teflon spacer are utilized. Fig. 6a shows the comparison of near-field responses of a meta-dimer with various dielectric surroundings. In experiments, we put various dielectric plates on the top surface of the meta-dimer. The measured near-field transmission spectra are shown in **Figure 6a**, when the meta-dimer is covered by various 1-mm-thick materials: air ($\varepsilon_r$ = 1.0), foam ($\varepsilon_r$ = 1.1), Teflon ($\varepsilon_r$ = 2.1), and RT5880 ($\varepsilon_r$ = 2.2). From the measured results, we observe that the invisibility dip (arising from interference between hexapole and quadrapole modes) has significant shift by covering different materials. The invisibility dips locate at 4.948 GHz, 4.876 GHz, 4.812 GHz, and 4.683 GHz, corresponding to air, foam, Teflon, and RT5880 respectively. That is to say, we obtained a 0.265 GHz (or 5.36%) shift for a 48.32% change in index (form air $n$ = 1 to RT5880 $n$ = 1.4832).

In addition to sensing different permittivity, the invisibility dips can also sense the same material of different thickness. Then we demonstrate this sensing performance of invisibility dips with Rogers RT5880 plates of different thickness (0.25 mm, 0.5 mm, 0.75 mm, 1.0 mm). Similar with above sensing experiment, the Rogers plates are placed on the top surface of the meta-dimer. We observe that in Figure 6b the invisibility dips shift from 4.788 GHz (with 0.25 mm Rogers plate) to the lower frequency 4.683 GHz (with a 1.0 mm Rogers plate).

## 7. Conclusion

In conclusion, we have observed strong-coupling-induced spectral dips in two vertically coupled identical SLSP resonators. Mode interferences induced by strong coupling are enhanced by gradually decreasing thickness of Teflon plates between the two SLSP resonators. Finally, mode crossing further verifies mode interference mechanism rather than intermode coupling, in which low-order modes show higher frequencies than high-order modes. Meanwhile, lifetime-contrast splittings are observed, where radiation coupling plays a significant role. The whole process is verified with transmission spectra and near-field imaging. Coupled mode theory has been developed to quantitatively describe the process. We also demonstrated the sensing capability of invisibility dips with different dielectric materials of the same thickness and same material with different thickness. Our studies may extend to surface plasmon systems in optical frequencies and open a new avenue for various subwavelength optical applications.

**Achnowledgements**

We thank Prof. John Pendry for kind suggestions. F. Gao, and Z. Gao contributed equally to this work. This work was partially supported by Nanyang Technological University for Start-up Grants, Singapore Ministry of Education under Grant No. MOE2015-T2-1-070 and Grant No. MOE2011-T3-1-005 and Grant No. MOE2015-T2-1-145, and the Program Grant (11235150003) from NTU-A*STAR Silicon Technologies Centre of Excellence.

[1]  A. Aubry, D. Lei, S. A. Maier, and J. B. Pendry, *Phys. Rev. Lett.* **2010**, 105, 233901.
[2]  U. Fano, *Phys. Rev.* **1961**, 124, 1866.
[3]  B. Luk'yanchuk, N. I. Zheludev, S. A. Maier, N. J. Halas, P. Nordlander, H. Giessen, and C. T. Chong, *Nature Mater.* **2010**, 9, 707.
[4]  A. E. Miroshnichenko, *Rev. Mod. Phys.* **2010**, 82, 2257.
[5]  Y. J. Bao, R. W. Peng, D. J. Shu, M. Wang, X. Lu, J. Shao, W. Lu, and N. B. Ming, *Phys. Rev. Lett.* **2008**, 101, 087401.
[6]  J. B. Pendry, D. Martin-Cano, and F. J. Garcia-Vidal, *Science*, **2004**, 305, 847-848.
[7]  F. J. Garcia-Vidal, L. Martin-Moreno, and J. B. Pendry, *J. Opt. A*, **2005**, 7, S97.

[8] A. Pors, E. Moreno, L. Martin-Moreno, J. B. Pendry, and F. J. Garcia-Vidal, *Phys. Rev. Lett.* **2012**, 108, 22.

[9] X. P. Shen, and T. J. Cui, *Laser Photonics Rev.* **2014**, 8, 137.

[10] P. A. Huidobro, X. Shen, J. Cuerda, E, Moreno, L. Martin-Moreno, F. J. Garcia-Vidal, T. J. Cui, and J. B. Pendry, *Phys. Rev. X*, **2014**, 4, 021003.

[11] F. Gao, Z. Gao, X. Shi, Z. Yang, X. Lin, and B. Zhang, *Opt. Express*, **2015**, 23, 6896.

[12] S. A. Maier, S. R. Andrews, L. Martin-Moreno, and F. J. Garcia-Vidal, *Phys. Rev. Lett.* **2006**, 97, 17.

[13] N. Yu, Q. J. Wang, M. A. Kats, J. A. Fan, S. P. Khanna, L. Li, A. G. Davies, E. H. Linfield, and F. Capasso, *Nature Mater.* **2010**, 9, 730-735.

[14] F. Gao, Z. Gao, X. Shi, Z. Yang, X. Lin, H. Xu, J. D. Joannopoulos, M. Soljacic, H. Chen, L. Lu, Y. Chong, and B. Zhang, *Nature Commun.* **2015**, 7, 11619.

[15] Z. Liao, B. C. Pan, X. Shen, and T. J. Cui, *Opt. Express*, **2014**, 22, 13.

[16] S. A. Maier, P. G. Kik, H. A. Atwater, S. Meltzer, E. Harel, B. E. Koel, and A. A. G. Requicha, *Nature Mater.* **2003**, 2, 229.

[17] B. Peng, S. K Ozdemir, W. Chen, F. Nori, and L. Yang, *Nature Commun.* **2014**, 5, 5082.

[18] R. H. Dicke, *Phys. Rev.* **1954**, 93, 99.

[19] H. Chen, J. Zhang, Y. Bai, Y. Luo, L. Ran, Q. Jiang, and J. A. Kong, *Opt. Express*, **2006**, 14, 12944.

[20] F. Gao, Z. Gao, Y. Zhang, X. Shi, Z. Yang, and B. Zhang, *Laser Photonics Rev.* **2015**, 9, 5.

[21] H. A. Haus, Waves and Fields in Optoelectronics, Prentice-Hall, Englewood Cliffs, NJ, **1984**.

[22] R. Taubert, M. Hentschel, J. Kastel, and Harald Giessen, *Nano Lett.* **2012**, 12, 1367.

[23] L. Verslegers, Z. Yu, Z. Ruan, P. B. Catrysse, and S. Fan, *Phys. Rev. Lett.* **2012**, 108, 083902.

[24] S. Zhang, Z. Ye, Y. Wang, Y. Park, G. Bartal, M. Mrejen, X. Yin, X. Zhang, *Phys. Rev. Lett.* **2012**, 109, 193902.

[25] J. B. Pendry, A. I. Fernandez-Dominguez, Y. Luo, and R. Zhao, *Nature Phys.* **2013**, 9, 518.

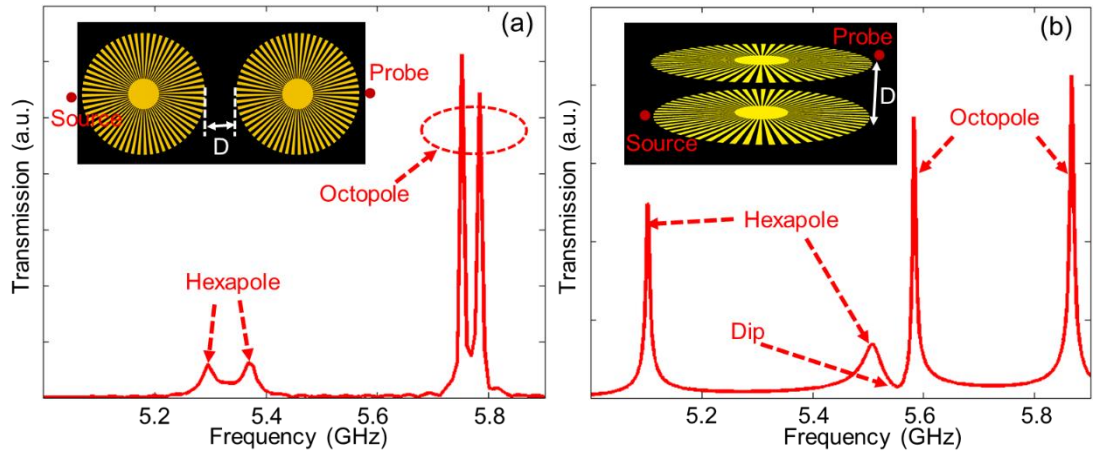

**Figure 1.** (a) Mode splitting in horizontally coupled SLSP resonators placed on a Teflon plated with thickness t = 6.1 mm. (b) Lifetime-contrast splitting and "invisibility dip" in vertically coupled resonators spaced with the same Teflon plate in (a).

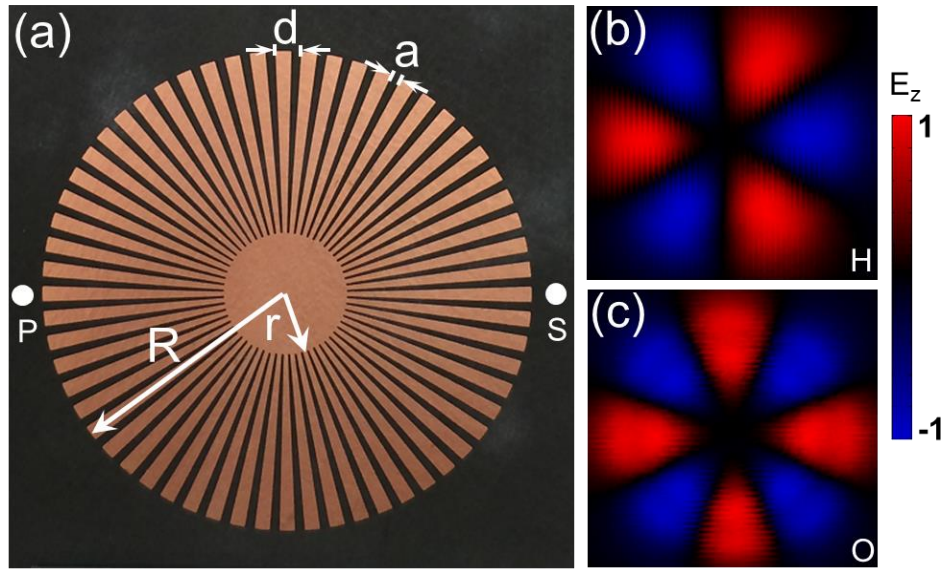

**Figure 2.** (a) Optical image of SLSP resonator. (b)-(c) Experimentally recorded $E_z$ field patterns on X-Y plane at 5.258 GHz, and 5.599 GHz respectively, which correspond to hexapole (H), and octopole (O) respectively.

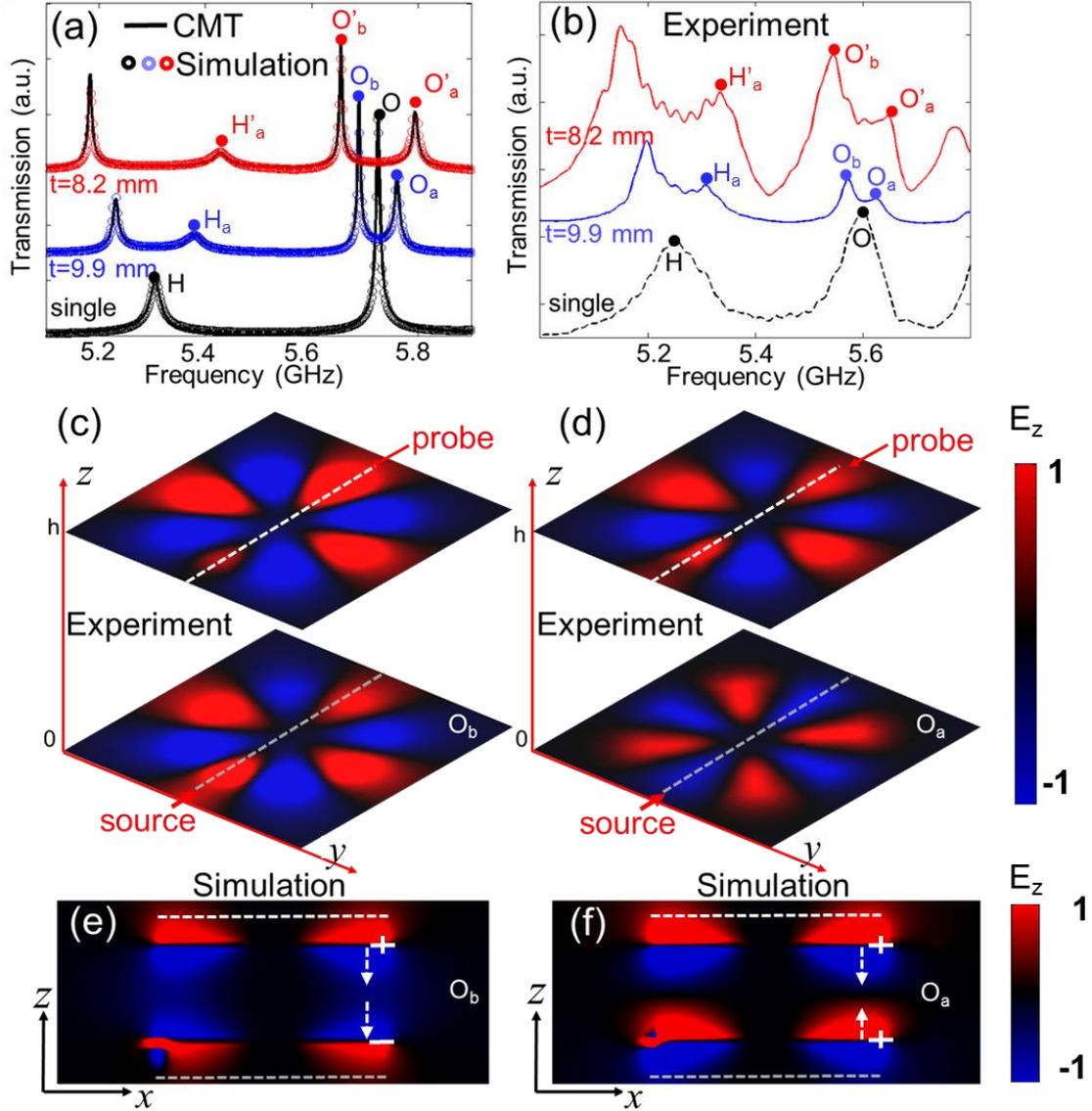

**Figure 3.** Lifetime-contrast splitting. (a) Simulated trasmission spectra of a single resonator (black circles), vertically coupled resonators with 9.9 mm-thick teflon spacer (blue cirlces), and with 8.2 mm-thick teflon spacer (red cirlces). Black solid curves are fitting results with CMT. (b) Detected transmission spectra of a single resonator (black dashed lines), 9.9 mm-thick teflon spacer (blue lines), and 8.2 mm-thick teflon spacer (red lines). (c)-(d) Experimentally recorded $E_z$ field patterns on X-Y plane at 5.58 GHz, and 5.62 GHz respectively, which correspond to octopole binding mode ($O_b$), and anti-binding mode ($O_a$) respectively. (e)-(f) Simulated $E_z$ field pattern on X-Z plane at $O_b$, and $O_a$ respectively, which correspond to (c) and (d) respectively.

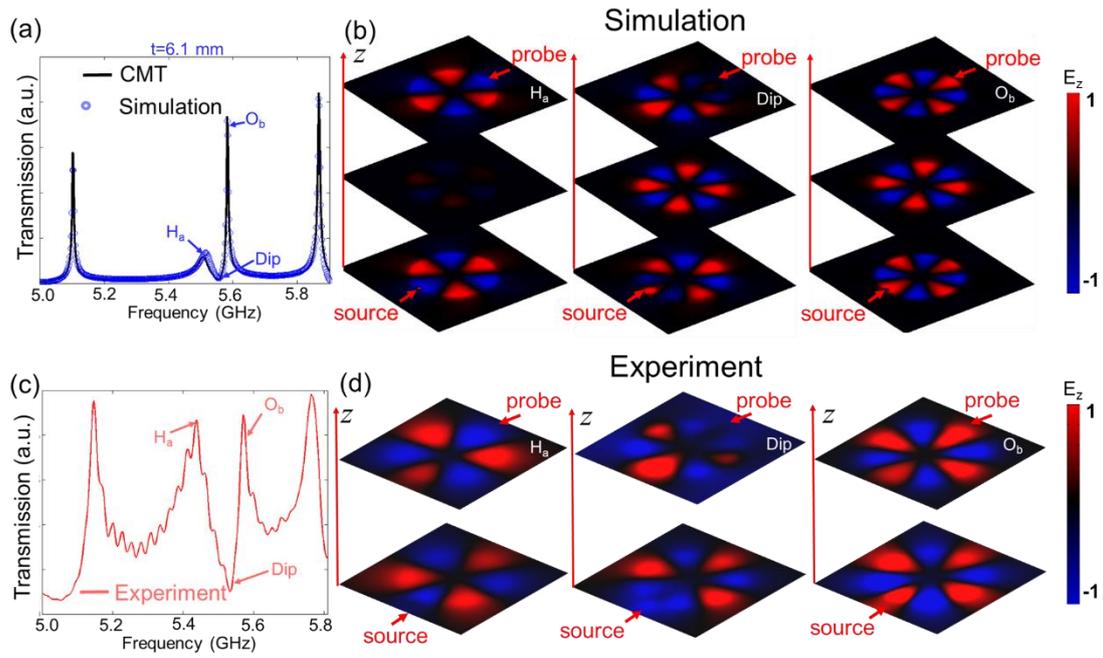

**Figure 4.** Strong-coupling-induced "invisibility dip". (a) Simulated transmission spectrum (blue circle line) with spacer thickness $t = 6.1$ mm, and fitting results by CMT (black solid line). (b) Simulated field patterns on top, middle (inside spacer), and bottom X-Y planes at three different frequencies ($H_a$, Dip, $O_b$) (c) Experimentally detected transmission spectrum (d) Experimentally recorded $E_z$ field patterns on top and bottom XY planes at $H_a$, Dip and $O_b$ respectively.

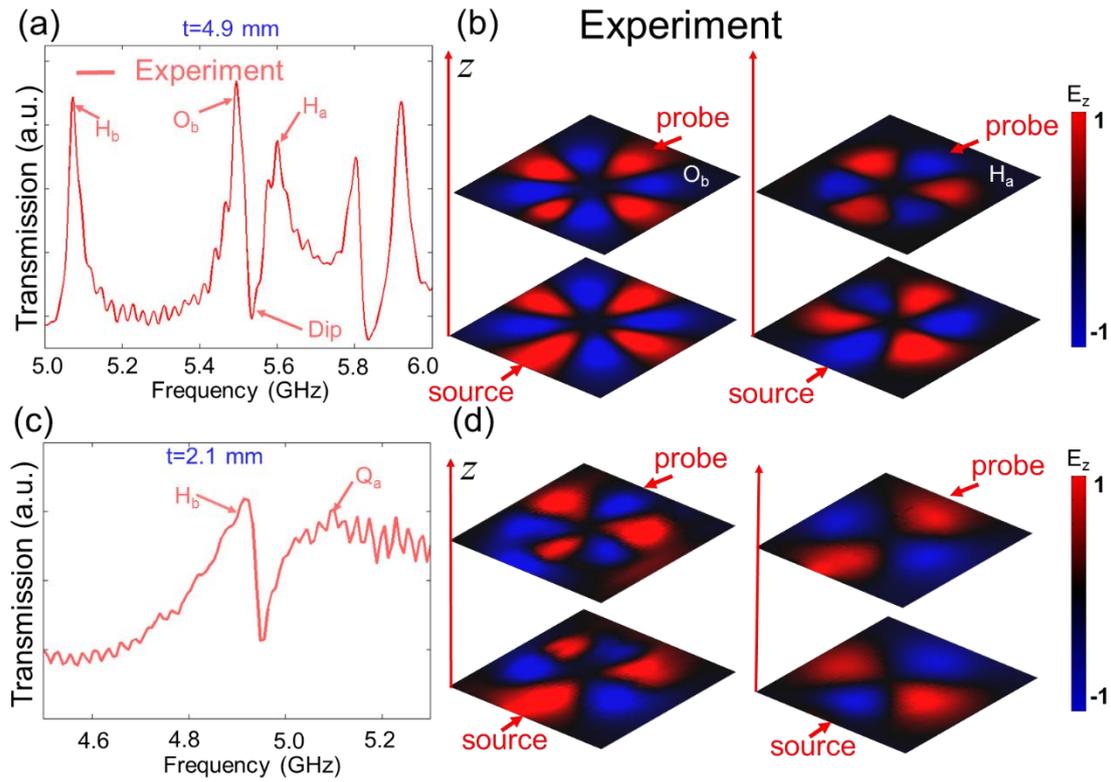

**Figure 5.** "Invisibility dip" and modes crossing. (a) Measured transmission spectrum with spacer thickness $t = 4.9$ mm. (b) Experimentally recorded $E_z$ field patterns on top and bottom XY planes at $H_a$, and $O_b$ respectively. (c) Measured transmission spectrum with spacer thickness $t = 2.1$ mm. (d) Experimentally recorded $E_z$ field patterns on top and bottom XY planes at $H_b$, and $Q_a$ respectively.

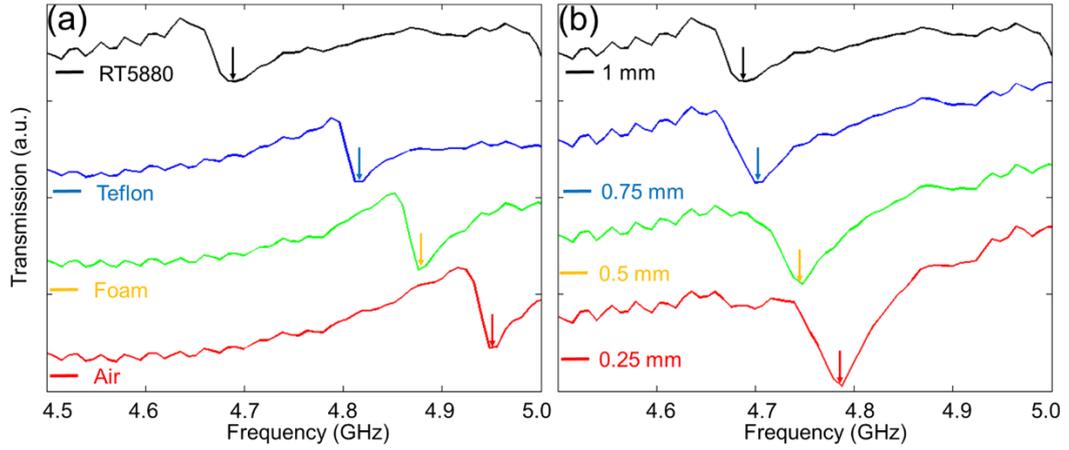

**Figure 6.** Sensing with invisibility dips. (a) Shift of invisibility dips by placing different dielectric plates on the top surface of the meta-dimer. (b) Shift of invisibility dips by placing Rogers RT5880 plates of different thickness on the top surface of the meta-dimer.